\documentclass[9pt,twocolumn,twoside]{opticajnl}
\journal{pr} % Choose journal (ao,jocn,josaa,josab,ol,optica,pr)

\setboolean{shortarticle}{false}
\usepackage{lineno}
\usepackage{dirtytalk}
\usepackage{siunitx}
\usepackage{braket}
\title{Pseudospin-2 in photonic chiral borophene}

\author[1,*]{Philip Menz}
\author[1]{Haissam Hanafi}
\author[2]{Daniel Leykam}
\author[1]{Jörg Imbrock}
\author[1,3]{Cornelia Denz}

\affil[1]{Institute of Applied Physics, University of Muenster, Muenster 48149, Germany}
\affil[2]{Centre for Quantum Technologies, National University of Singapore, 3 Science Drive 2, 117543, Singapore}
\affil[3]{Physikalisch-Technische Bundesanstalt (PTB), Bundesallee 100, 38116 Braunschweig, Germany}

\affil[*]{Corresponding author: philip.menz@uni-muenster.de}

\begin{abstract}
  Pseudospin is an angular momentum degree of freedom introduced in analogy to the real electron spin in the effective massless Dirac-like equation used to describe wave evolution at conical intersections such as the Dirac cones of graphene. Here, we study a photonic implementation of a chiral borophene allotrope hosting a pseudospin-2 conical intersection in its energy-momentum spectrum. The presence of this fivefold spectral degeneracy gives rise to quasiparticles with pseudospin up to $\pm2$. We report on conical diffraction and pseudospin-orbit interaction of light in photonic chiral borophene, which, as a result of topological charge conversion, leads to the generation of highly charged optical phase vortices.
\end{abstract}

\setboolean{displaycopyright}{false}

\begin{document}

\maketitle

\section{Introduction}
Conical intersections are features of parameter spaces where two or more energy surfaces become degenerate at one point, while staying linear in its vicinity. Two prominent examples are Hamilton's diabolical point in biaxial crystals in optics~\cite{Berry2007} and Dirac cones in solid state's iconic material graphene~\cite{Geim2007}. For the latter, due to the mathematical analogy with the Dirac equation for massless electrons, a microscopic degree of freedom called pseudospin was introduced~\cite{Mecklenburg2011}. Unlike the polarization-related photon spin or the intrinsic spin of electrons, this form of angular momentum is not associated with any intrinsic property of particles. Instead, it arises from the substructure of space given by the periodic potential in which the wave function resides~\cite{Leykam2016}.

Psuedospin quasiparticles in periodic lattices with conical intersections represent a practical test bed for observing quantum relativistic effects implied by the Dirac equation and its higher-spin versions like Klein tunneling~\cite{Katsnelson2006b,Urban2011} or Zitterbewegung~\cite{Katsnelson2006a}. Photonic model systems such as evanescently coupled waveguides, so-called photonic lattices, allow observing a variety of classical analogs of both relativistic and non-relativistic quantum phenomena associated with the evolution of electrons in periodic potentials~\cite{Longhi2009,Dreisow2010a} thanks to the formal correspondence between the Schrödinger equation and the paraxial wave equation. A convenient feature of photonic lattices is that they provide direct access to the evolution of the wave function during propagation. Therefore, a natural step was to use a photonic platform to realize periodic lattices hosting conical intersections in their spectrum and to demonstrate their peculiarities by studying light propagation through them. This has already led to realizations of pseudospin-1/2 photonic graphene~\cite{Song2015}, the pseudospin-1 photonic Lieb lattice~\cite{Diebel2016b} and the Lieb-kagome transition lattice~\cite{lang2023tilted}.

A challenging open problem in artificial lattice systems is the design of conical intersections with higher pseudospin values~\cite{Leykam2016}. Although there have been proposals for generalized conical intersections with arbitrary pseudospin~\cite{Dora2011,Lan2011}, and pseudospin-2 ones have been considered theoretically~\cite{feng2019effective,kim2022mode}, no realistic system containing a conical intersection with pseudospin higher than one has been demonstrated until now.

Here we present a photonic chiral borophene lattice hosting a pseudospin-2 conical intersection in its band structure at the center of its Brillouin zone. We derive the five pseudospin eigenstates using both an intuitive and a rigorous mathematical-analytical approach and numerically study their conical diffraction during propagation through the photonic lattice. We prove the interaction of pseudospin and orbital angular momentum by directly observing topological charge conversion giving rise to optical phase vortices in the conically diffracted output light fields. Here, with topological charge we mean the winding number of the optical wavefront around the vortex core. Our results apply to various other wave systems beyond photonics such as metamaterials~\cite{Nakata2012a}, Bose-Einstein and polariton condensates~\cite{Taie2015,Baboux2016}, and importantly also to electronic wave functions in atomic borophene allotropes.

\section{Results}
\subsection{Photonic chiral borophene and its pseudospin-2 conical intersection}
\begin{figure}
\centering\includegraphics[width=\linewidth]{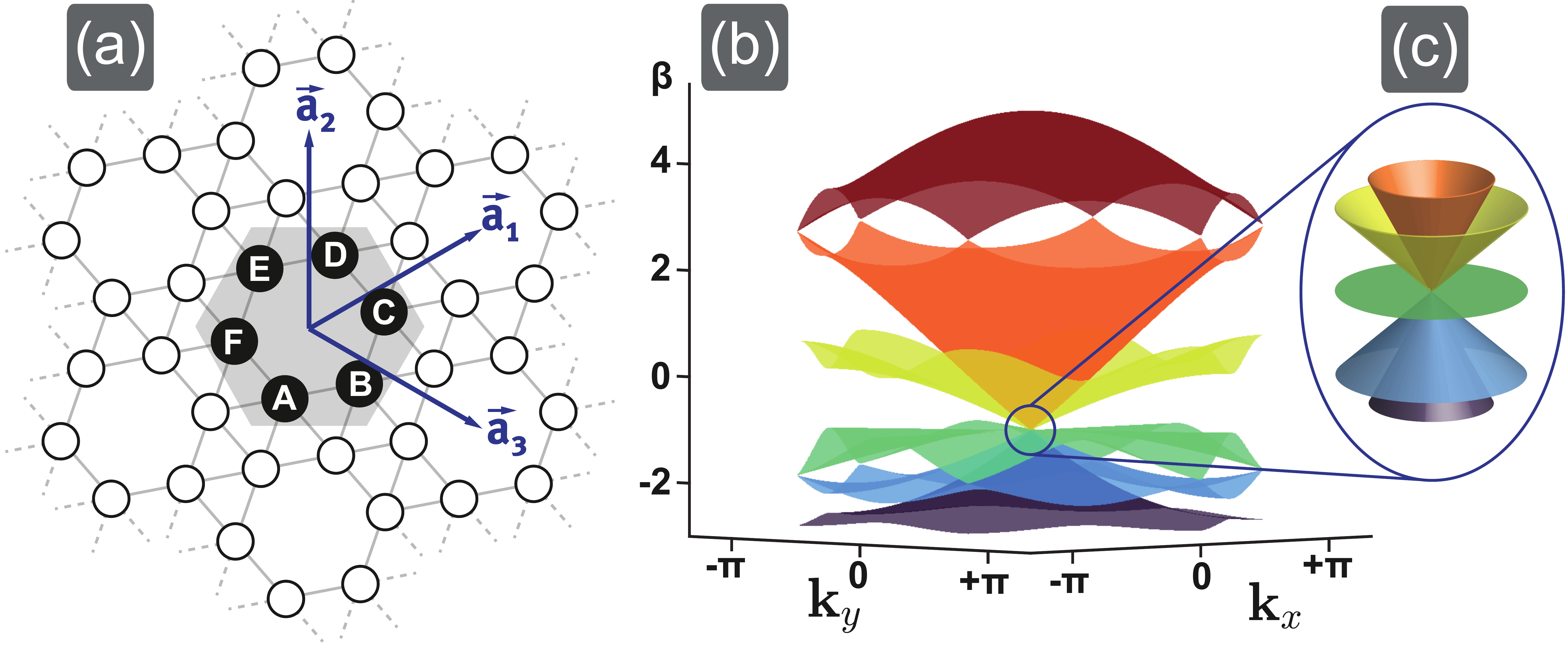} 
\caption{The chiral borophene lattice and its band structure. (a) Schematic of the lattice with the unit cell in gray and the lattice vectors $\mathbf{a_1}$, $\mathbf{a_2}$, and $\mathbf{a_3}$. The lattice sites are labeled from A to F. (b) Tight-binding band structure in the hexagonal Brillouin zone, calculated using $d=t=1$ for nearest neighbors only. (c) Zoomed-in view of the linear dispersion close to the degenerate point showing an idealized pseudospin-2 conical intersection.}
\label{fig:latt_overview}
\end{figure}

Fig.~\ref{fig:latt_overview}a shows a sketch of the chiral borophene lattice. The lattice has a hexagonal unit cell with six lattice sites labeled A to F. This configuration has been calculated to be stable as a planar sheet of boron atoms~\cite{Zope2011,Yi2017}. This distinguishing feature may lead to the realization of pseudospin-2 conical intersections in a solid-state 2D material. The geometry of the lattice corresponds to an Archimedean tiling of the plane. More precisely, it is the $(3^4,6)$ or snub hexagonal tiling which interestingly exists in two chiral variants~\cite{Grunbaum1977}. The band structure of the chiral borophene lattice is depicted in Fig.~\ref{fig:latt_overview}b. Intriguing features include the pseudospin-1/2 Dirac cones~\cite{CrastodeLima2019a} and the partially flat band~\cite{menz2022noncontractible}. The peculiarity we are interested in this work is the conical intersection of five bands at the degenerate $\Gamma$-point shown in the zoom-in of Fig.~\ref{fig:latt_overview}c. This fivefold degeneracy has been shown to be protected by site-permutation symmetries and therefore be robust to long-range isotropic interactions such as $\text{p}^\text{th}$ nearest neighbor hopping for $\text{p}\rightarrow \infty $~\cite{CrastodeLima2020}.

An experimental photonic lattice realization of chiral borophene would rely on an array of evanescently coupled waveguides arranged according to the lattice geometry shown in Fig.~\ref{fig:latt_overview}a. The waveguides could either be created by femtosecond direct laser writing~\cite{szameit2010discrete} or by optical induction in a photorefractive medium~\cite{Song2015}. In this study, we use a numerical approach which has proven to agree very well with experimental realizations~\cite{Diebel2016b}.
In our photonic waveguide model, we apply a tight-binding approximation as a discrete model describing the evanescent coupling between the lattice sites. Considering only nearest neighbor coupling, we obtain the following $k$-space ($\mathbf{k}=(k_x,k_y)$) Hamiltonian
\begin{equation}\label{eq:Ham_snub_boro}
    \hat{H}_\mathbf{k} = t \begin{pmatrix}
0 & 1 & e^{-\text{i}\mathbf{a}_1\mathbf{k}} & e^{-\text{i}\mathbf{a}_2\mathbf{k}} & e^{-\text{i}\mathbf{a}_2\mathbf{k}} & 1 \\
1 & 0 & 1 & e^{-\text{i}\mathbf{a}_2\mathbf{k}} & e^{\text{i}\mathbf{a}_3\mathbf{k}} & e^{\text{i}\mathbf{a}_3\mathbf{k}} \\
e^{\text{i}\mathbf{a}_1\mathbf{k}} & 1 & 0 & 1 & e^{\text{i}\mathbf{a}_3\mathbf{k}} & e^{\text{i}\mathbf{a}_1\mathbf{k}}\\
e^{\text{i}\mathbf{a}_2\mathbf{k}} & e^{\text{i}\mathbf{a}_2\mathbf{k}} & 1 & 0 & 1 & e^{\text{i}\mathbf{a}_1\mathbf{k}}\\
e^{\text{i}\mathbf{a}_2\mathbf{k}} & e^{-\text{i}\mathbf{a}_3\mathbf{k}} & e^{-\text{i}\mathbf{a}_3\mathbf{k}} & 1 & 0 & 1\\
1 & e^{-\text{i}\mathbf{a}_3\mathbf{k}} & e^{-\text{i}\mathbf{a}_1\mathbf{k}} & e^{-\text{i}\mathbf{a}_1\mathbf{k}} & 1 & 0
\end{pmatrix},
\end{equation}
where the lattice vectors are given by $\mathbf{a_1} = d/2(\sqrt{3},1)$, $\mathbf{a_2} = d(0,1)$, and $\mathbf{a_3} = \mathbf{a_1} - \mathbf{a_2}$, $d$ is the lattice constant and $t$ is the coupling strength. Setting $d=t=1$ without restriction of generality, the eigenvalues of $\hat{H}$ give the spectrum $\beta(\mathbf{k})$ shown in Fig.~\ref{fig:latt_overview}b. In our case, the band structure with its propagation constants $\beta(\mathbf{k})$ (which describes the rate of phase evolution in the propagation
direction) represents a diffraction relation describing the \textit{spatial} evolution dynamics of photonic wave functions in the lattice. In an atomic borophene lattice, this corresponds to an energy spectrum describing the \textit{temporal} evolution of the electronic wave function.
\begin{figure}
\centering\includegraphics[width=\linewidth]{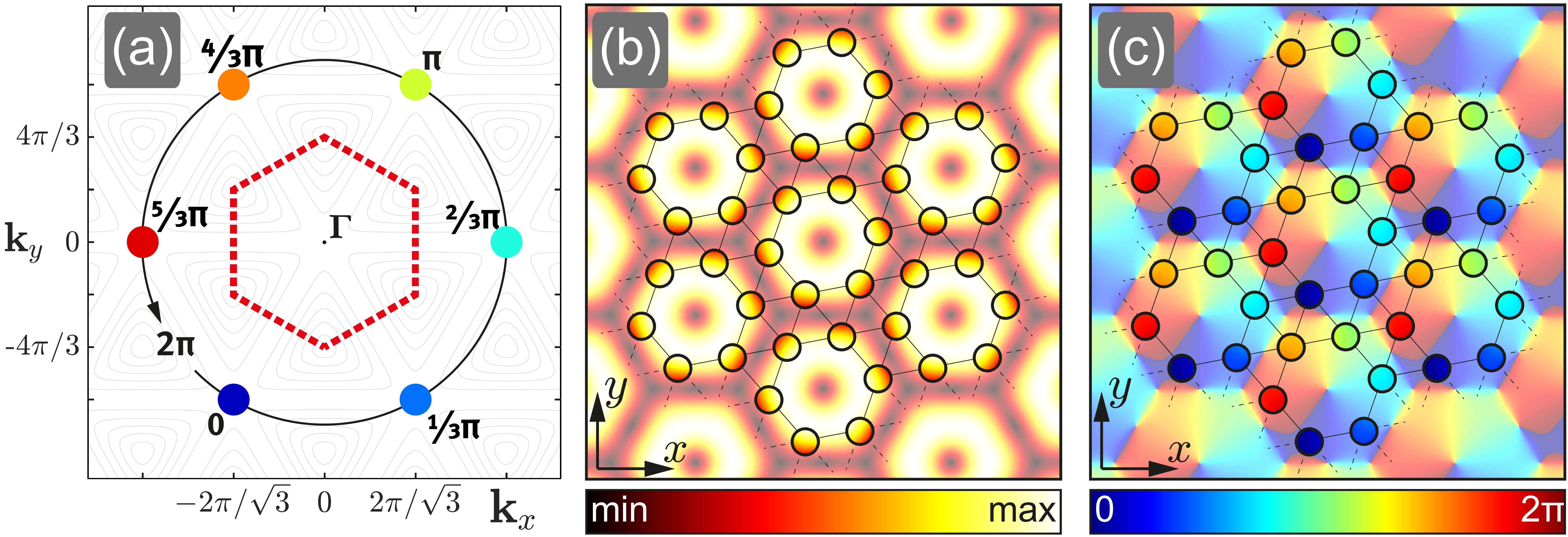} 
\caption{Derivation of pseudospin eigenstate $\ket{\psi_{-2}}$. (a) Six plane waves in k-space at the $\Gamma$-points surrounding the first Brillouin zone (dashed red line) forming a hexagonal discrete vortex with $l=+1$. The shown isolines of the third (green) band help to identify the exact k-space excitation. (b) and (c) Real space transverse amplitude and phase profile of the resulting discrete nondiffracting beam with overlaid chiral borophene lattice.}
\label{fig:deriv}
\end{figure}

At the singular $\Gamma$-point, a single-band approximation fails. It is, however, possible to understand complex multi-band effects by introducing the pseudospin as an analog to a real spin. To obtain the five pseudospin eigenstates describing the conical intersection of the chiral borophene lattice, we can proceed analytically and Taylor expand $\hat{H}$ around the singular point~\cite{Leykam2012}. A detailed analytic derivation is presented in \ref{sec:anal_deriv}. Here, we showcase how to derive the pseudospin eigenstates intuitively.

\begin{figure}
\centering\includegraphics[width=\linewidth]{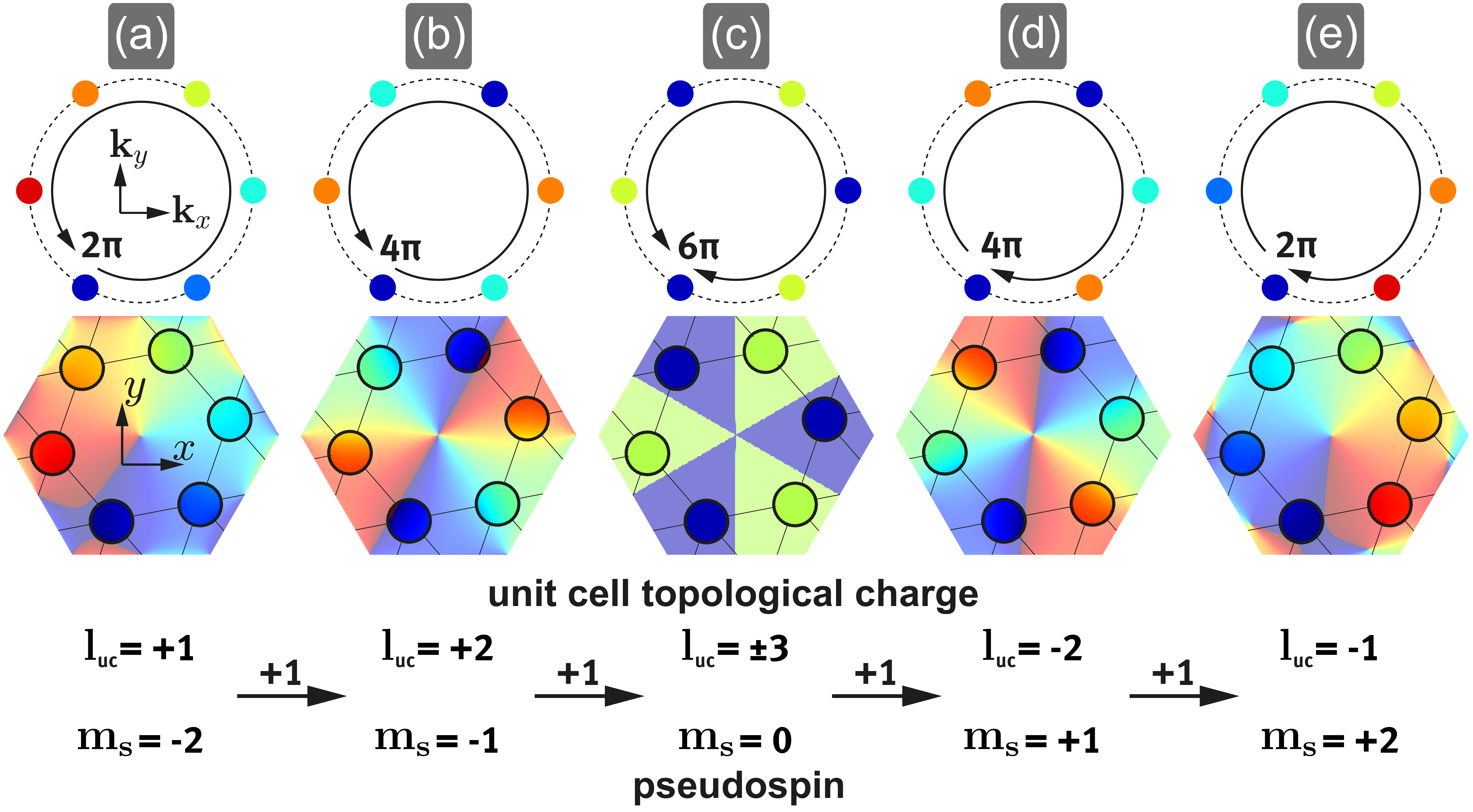} 
\caption{Order of the five pseudospin states for $\mathbf{m_s}$ increasing from $\mathbf{m_s=-2}$ on the left to $\mathbf{m_s=+2}$ on the right. The top row shows the discrete vorticity in k-space. The bottom row depicts the vortices of the corresponding discrete nondiffracting fields in real space , and therefore of the pseudospin eigenstates in the unit cell. The k-space and real space topological charge increases by unity from left to right. This is also reflected in the pseudospin value.}
\label{fig:ps_ranking}
\end{figure}

We start by exciting six $\Gamma$-points at the centers of the six Brillouin zones surrounding the first one. The resulting interference of six plane waves is known to give rise to a family of discrete nondiffracting beams~\cite{Boguslawski2011a}. For six plane waves with a specific phase relations, resembling a discrete phase vortex, the nondiffracting fields are periodic with a sixfold symmetry. Five of these cases lead to the desired pseudospin eigenstates as illustrated for the pseudospin eigenstate with $\ket{\psi_{m_s=-2}}$ in Fig.~\ref{fig:deriv}. For differently charged discrete phase vortices of the six plane waves we obtain the other four eigenstates (see \ref{sec:deriv}). In the sublattice basis the normalized eigenstates finally read
\begin{equation}\label{eq:eigenstates}
\begin{split}
    \ket{\psi_{-2}} &= \frac{1}{\sqrt{6}} \begin{pmatrix} +1 & e^{\text{i}\frac{1}{3}\pi} & e^{\text{i}\frac{2}{3}\pi} & -1 & e^{\text{i}\frac{4}{3}\pi} & e^{\text{i}\frac{5}{3}\pi} \end{pmatrix}^T,\\
    \ket{\psi_{-1}} &= \frac{1}{\sqrt{6}} \begin{pmatrix} +1 & e^{\text{i}\frac{2}{3}\pi} & e^{\text{i}\frac{4}{3}\pi} & +1 & e^{\text{i}\frac{2}{3}\pi} & e^{\text{i}\frac{4}{3}\pi} \end{pmatrix}^T,\\
    \ket{\psi_{0}} &= \frac{1}{\sqrt{6}} \begin{pmatrix} +1 & -1 & +1 & -1 & +1 & -1 \end{pmatrix}^T,\\
    \ket{\psi_{+1}} &= \frac{1}{\sqrt{6}} \begin{pmatrix} +1 & e^{\text{i}\frac{4}{3}\pi} & e^{\text{i}\frac{2}{3}\pi} & +1 & e^{\text{i}\frac{4}{3}\pi} & e^{\text{i}\frac{2}{3}\pi} \end{pmatrix}^T,\\
    \ket{\psi_{+2}} &= \frac{1}{\sqrt{6}} \begin{pmatrix} +1 & e^{\text{i}\frac{5}{3}\pi} & e^{\text{i}\frac{4}{3}\pi} & -1 & e^{\text{i}\frac{2}{3}\pi} & e^{\text{i}\frac{1}{3}\pi} \end{pmatrix}^T,
\end{split}
\end{equation}
where the six entries represent the complex amplitudes at lattice sites A to F. $\ket{\psi_{-2,-1,0,+1,+2}}$ are all eigenvectors of $\hat{H}(0,0)$ for the eigenvalue $\beta=-1$. Together with the eigenvector of the sixth band, wich has a different eigenvalue or propagation constant of $\beta=5$, $\ket{\psi_{\beta=5}}=
\begin{pmatrix}
1 & 1 & \cdots & 1
\end{pmatrix}^T$,
they form an orthonormal orbital angular momentum basis for the unit cell. Our intuitive derivation allows obtaining the five pseudospin eigenstates, although what is still missing is the assignment to the correct pseudospin value ranging from $m_s=-2$ to $m_s=+2$. We obtain this directly from the analytical derivation presented in \ref{sec:anal_deriv}. However, again an intuitive explanation arises from comparing the phase distributions of the different pseudospin eigenstates. As can be seen in Fig.~\ref{fig:ps_ranking}, the order relies on a difference in the topological charges of the \textit{microscopic} optical phase vortices internal to the unit cell $l_{\text{uc}}$: when we increase the pseudospin by unity from $\ket{\psi_{-2}}$ to $\ket{\psi_{-1}}$, the vorticity is also increased from $l_{\text{uc}}=+1$ to $l_{\text{uc}}=+2$. The same happens when going from $\ket{\psi_{-1}}$ to $\ket{\psi_{0}}$ with $l_{\text{uc}}=+2$ increasing to $l_{\text{uc}}=+3$. $\ket{\psi_{0}}$ plays a crucial role in this process as the discreteness of its internal topological charge can be regarded to have either a positive or a negative value of $l_{\text{uc}}=\pm3$. As a result, from $\ket{\psi_{0}}$ to $\ket{\psi_{+1}}$ we again have a unitary increase from $l_{\text{uc}}=-3$ to $l_{\text{uc}}=-2$. For the last transition from $\ket{\psi_{+1}}$ to $\ket{\psi_{+2}}$, the relation is once again valid changing from $l_{\text{uc}}=-2$ to $l_{\text{uc}}=-1$. At first glance this ordering may appear a bit counterintuitive. In particular one might ask why there is no 1:1 correspondence between pseudospin and the internal topological charge in the form $m_s = l_{\text{uc}}$. The answer is that the state having $l_{\text{uc}}=0$ has a propagation constant $\beta\neq -1$ and due to it not being part of the conical intersection, it does not couple to the other states. This prevents the transition $l_{\text{uc}}= -1 \rightarrow l_{\text{uc}}=0 \rightarrow l_{\text{uc}}= +1$ which would be necessary to go from $\ket{\psi_{-1}}$, over $\ket{\psi_{0}}$ to $\ket{\psi_{+1}}$ in case of a direct correspondence between pseudospin and internal topological charge.

\subsection{Conical diffraction and topological charge conversion}
To confirm that chiral borophene indeed hosts a pseudospin-2 conical intersection in its band structure and to validate the derived pseudospin eigenstates, we perform numerical experiments of light propagation in the lattice (see \ref{sec:methods} for details on the numerical methods). The simulations are based on the paraxial wave equation and are carried out via a standard pseudo-spectral split-step propagation method~\cite{Sharma2004a}. The numerical parameters are chosen to be within the experimental reach and match those in previously reported experiments in laser-written photonic lattices with a refractive index contrast of the waveguides $\Delta n = 1.3 \times 10^{-3}$, a wavelength of $\lambda = \SI{532}{\nano\metre}$, and a nearest neighbor waveguide separation of $\Lambda = \SI{18}{\micro\metre}$~\cite{hanafi2022localized,hanafi2022b}.

\begin{figure*}
\centering\includegraphics[width=\linewidth]{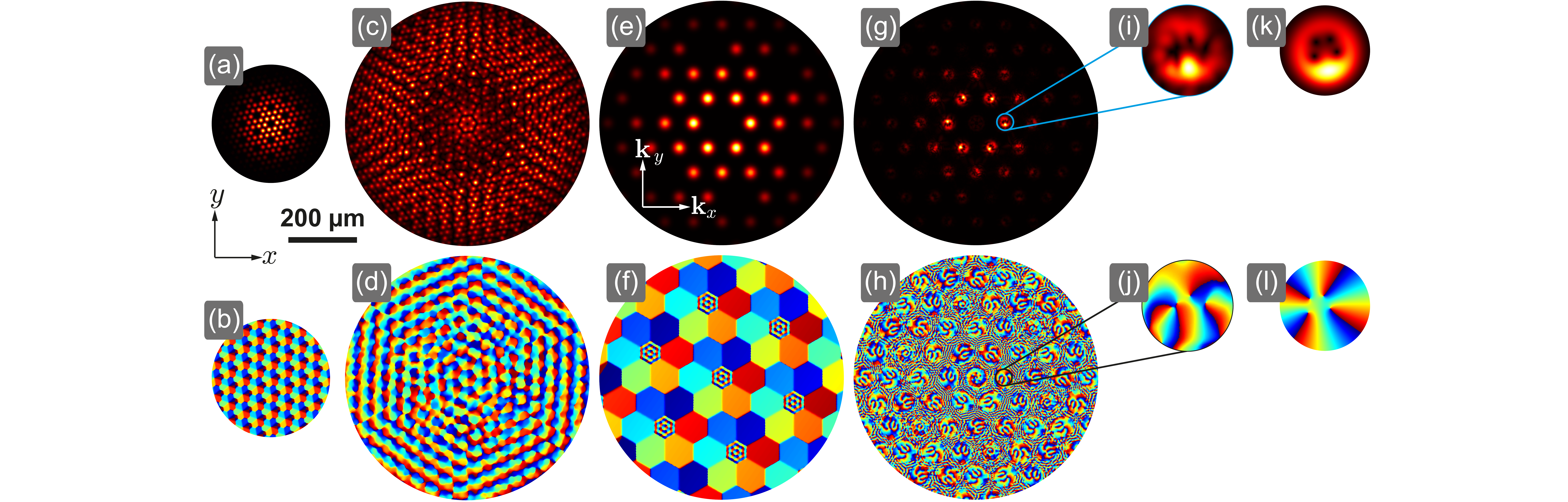}
\caption{Numerical simulation of conical diffraction and pseudospin-mediated vortex generation in photonic borophene. (a)-(b) Amplitude and phase of the input light field given by the pseudospin state $m_s=-2$ multiplied by a Gaussian envelope. (c)-(d) Output after propagation in the lattice. (e)-(f) Same as (a)-(b), but in Fourier-space. (g)-(h) Same as (c)-(d), but in Fourier-space. (i)-(j) Zoom-in showing one of the spectral components. (k)-(l) Ideal linear superposition of Laguerre-Gaussian modes $\text{LG}_{0,0} + \text{LG}_{0,-1} + \text{LG}_{0,-2} + \text{LG}_{0,-3} + \text{LG}_{0,-4}$ for comparison.}
\label{fig:output_real_k}
\end{figure*}

We excite the lattice with a light field given by the pseudospin eigenstate $\ket{\psi_{-2}}$ multiplied by a Gaussian envelope with $\text{FWHM} = \SI{120}{\micro\metre}$, as shown in Fig.~\ref{fig:output_real_k}a-b. In k-space, this corresponds to a Gaussian instead of a point-like excitation at the conical intersection (Fig.~\ref{fig:output_real_k}e-f). After propagation in the lattice for the distance $z=\SI{7.12}{\centi\metre}$ which corresponds to four coupling lengths $L_c$ for the chosen parameters, we clearly identify the conical diffraction in the output field as shown in Fig.~\ref{fig:output_real_k}c-d. Exactly at the $\Gamma$-point the Bloch bands become degenerate, and thus a plane wave input state with an arbitrary pseudospin would be invariant under propagation. However, a finite-size input beam will excite a range of wavevectors in the vicinity of the $\Gamma$-point. Away from the $\Gamma$-point, the degeneracy is lifted and thus, in general, eigenstates of the pseudospin will not coincide with the Bloch wave eigenstates of the Hamiltonian. Therefore there will be a coupling between different pseudospin eigenstates during propagation at a rate determined by the splitting of the Bloch waves' propagation constants. For this coupling between different pseudospin eigenstates to take place, there has to be a compensation for the difference in their pseudospin values based on the conservation of total angular momentum. This compensation leads to topological charge conversion in the form of optical vortices in the \textit{macroscopic} phase profiles of the output fields for the different pseudospin eigenstates. If we excite the lattice with a pseudospin state $m_{s}^{\text{in}}$, the topological charge of the vortices present in the decomposed output fields with pseudospin $m_{s}^{\text{out}}$  follow the relation
\begin{equation}
    l = m_{s}^{\text{in}}-m_{s}^{\text{out}}.
\end{equation}
Accordingly, the output field in Fig.~\ref{fig:output_real_k}c-d is composed of a superposition of $\ket{\psi_{-2,-1,0,+1,+2}}$ with phase vortices of topological charge $l=0,-1,-2,-3,-4$, respectively. In order to confirm this hypothesis, we look at the spectral components by performing a Fourier transform to the fields. In the Fourier transform we can see multiple spectral components at the centers of the higher-order Brillouin zones. While the higher components come from the waveguide structure and the degree of localization of the waveguide modes, we are interested in the symmetry properties which are fully captured by the six components at the centers of the second Brillouin zone. While the input (Fig.~\ref{fig:output_real_k}e-f) is composed of Gaussian spots at the center of the Brillouin zones, the output (Fig.~\ref{fig:output_real_k}g-h) is composed of more complex spots which we can identify as a superposition of Laguerre-Gaussian modes $\text{LG}_{0,0} + \text{LG}_{0,-1} + \text{LG}_{0,-2} + \text{LG}_{0,-3} + \text{LG}_{0,-4}$.
This is clearly seen by comparing one of the spectral components in the output (Fig.~\ref{fig:output_real_k}i-j) with an ideal superposition of the Laguerre-Gaussian modes (Fig.~\ref{fig:output_real_k}k-l). Crucially, both fields display a quadruply-charged optical phase vortex peculiar for a pseudospin-2 conical intersection resulting from the conservation of angular momentum going from the pseudospin state $\ket{\psi_{-2}}$ in the input to $\ket{\psi_{+2}}$ in the output.

\begin{figure}
\centering\includegraphics[width=\linewidth]{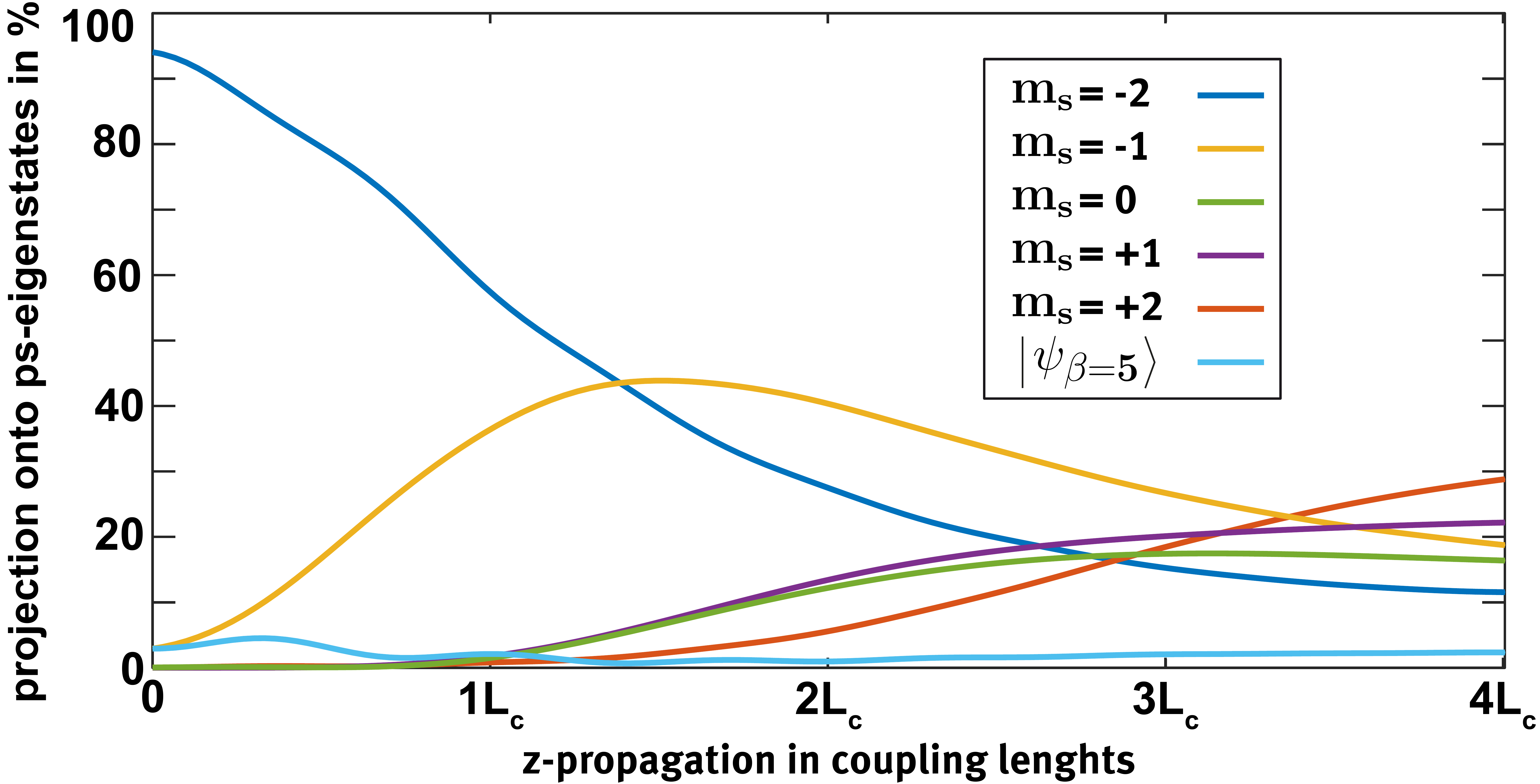} 
\caption{Projection onto pseudospin eigenstates during propagation and conical diffraction of input state $\ket{\psi_{-2}}$. Percentages of the different pseudospin eigenstates in the total field versus propagation distance in coupling lengths $L_c$ in the photonic lattice.}
\label{fig:percentages}
\end{figure}

These results already confirm the central message of this work: in demonstrating the existence of a pseudospin-2 conical intersection with five conically diffracting eigenstates in the linear spectrum of a chiral borophene lattice and the generation of highly charged optical vortices. Going beyond the spectral analysis, we give in the following a more detailed picture of the propagation dynamics in our photonic lattice close to the spectral singular point. To this aim, we decompose the output light field into the respective pseudospin components. This significantly simplifies the output phase profiles and allows the underlying mechanisms to be better elucidated. We carry out the decomposition by projecting the output field shown in Fig.~\ref{fig:output_real_k}c-d, unit cell by unit cell, onto the pseudospin eigenstates of Eq.~\ref{eq:eigenstates}~\cite{Diebel2016b}. For each unit cell we obtain five complex values representing the amplitude and phase of the respective pseudospin eigenstate (see \ref{sec:methods} for details).
\begin{figure}
\centering\includegraphics[width=\linewidth]{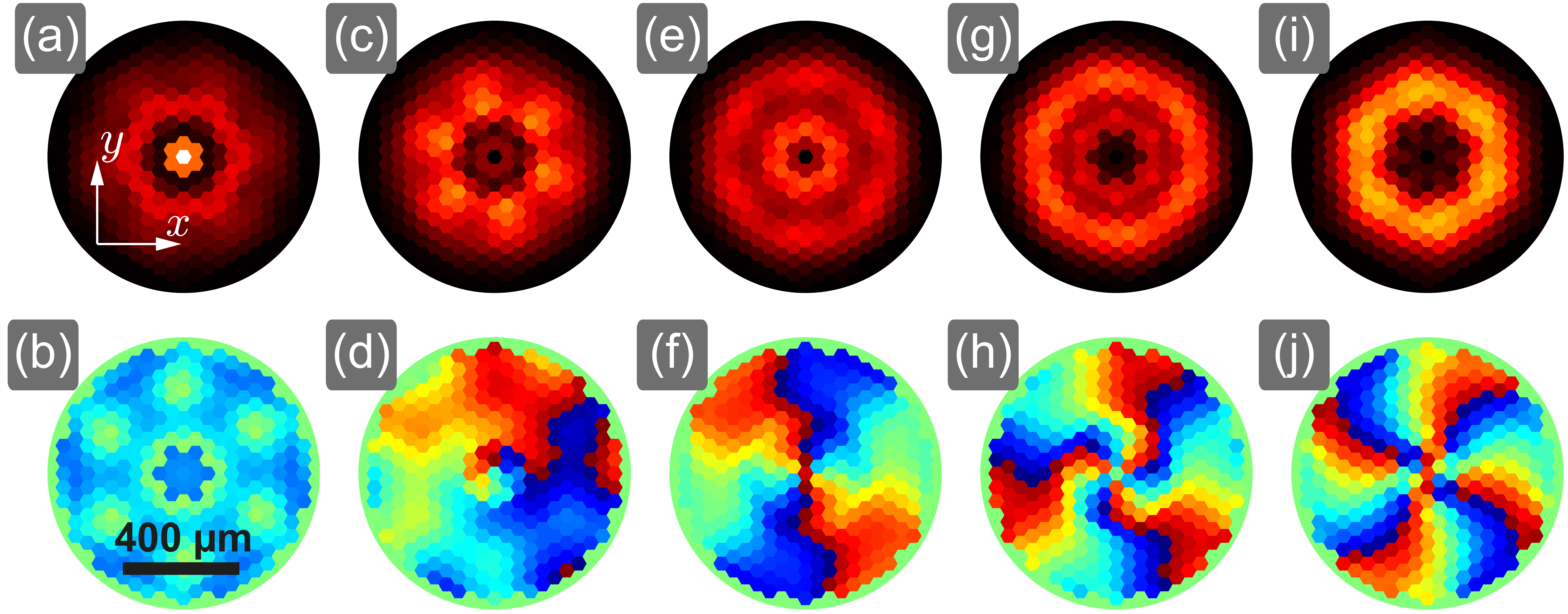} 
\caption{Projection of the conical diffraction output field onto the five pseudospin eigenstates. Each hexagonal pixel represents one unit cell. (a)-(b) Amplitude and phase of the projection onto $\ket{\psi_{-2}}$; (c)-(d) onto $\ket{\psi_{-1}}$; (e)-(f) onto $\ket{\psi_{0}}$; (g)-(h) onto $\ket{\psi_{+1}}$; (i)-(j) onto $\ket{\psi_{+2}}$.}
\label{fig:output_decomposed}
\end{figure}
We represent those values as hexagonal pixels in Fig.~\ref{fig:output_decomposed} for the $\ket{\psi_{-2}}$ input (for the remaining cases see \ref{sec:remaining}). In the phase profiles of the projections, we obtain optical phase vortices with topological charges following the relation $l = m_{s}^{\text{in}}-m_{s}^{\text{out}}$. Of particular interest is the $l=-4$ phase vortex that arises when projecting onto the pseudospin state $\ket{\psi_{+2}}$, since it is characteristic for a pseudospin-2 conical intersection. The phase vortices appear due to conservation of total angular momentum as the pseudospin value increases stepwise from $m_s=-2$ to $m_s=+2$ during propagation. To reveal the dynamics of this process we numerically solve the beam propagation in the photonic lattice according to the coupled differential equations of a discrete tight-binding model. As shown further in \ref{sec:comparison}, the tight-binding simulations match the continuous model ones extremely well. We then decompose the output field during propagation in the chiral borophene photonic lattice for different $z$-values. For each step, we calculate the probability amplitude of the total field for each pseudospin eigenstate and for the eigenstate of the sixth band $\ket{\psi_{\beta=5}}$. Thus, we obtain curves of projection percentages with respect to the z-propagation as depicted in Fig.~\ref{fig:percentages}.
At $z=0$, the lattice is excited with a light field primarily in the $\ket{\psi_{-2}}$-state. There are also minor components in $\ket{\psi_{-1}}$ and $\ket{\psi_{\beta=5}}$ due to the finite size of the Gaussian envelope. During propagation, we observe the share of the field being in $\ket{\psi_{-2}}$ to decrease, while sequentially the shares in $\ket{\psi_{-1}}$, $\ket{\psi_{0}}$, $\ket{\psi_{+1}}$, and $\ket{\psi_{+2}}$ increase. This shows that the pseudospin gradually increases in the order presented in Fig.~\ref{fig:ps_ranking} as it converts from $m_s=-2$ to $m_s=+2$, and that there is no coupling via the state of the sixth band $\ket{\psi_{\beta=5}}$ with $l_{\text{uc}}=0$. This clearly confirms that phase vortices form to compensate for the difference in internal topological charge $l_{\text{uc}}$ between the different pseudospin eigenstates. In this picture, the explanation of pseudospin-orbit interaction appears natural. Both are forms of orbital angular momentum of light, the first is \textit{microscopic} and internal to the unit cell while the latter is \textit{macroscopic} in the form of optical phase vortices in the total conically diffracted output field.

\section{Discussion}
In conclusion, we have studied a novel type of pseudospin-2 states in a photonic chiral borophene lattice at its fivefold conical intersection. We have numerically studied conical diffraction with topological charge conversion leading to the formation of optical phase vortices with topological charge values as high as $l=\pm4$. We are able to unveil this conversion as being the result of pseudospin-orbit interaction and conservation of total angular momentum. Moreover, it has been shown that the underlying mechanism is of topological origin due to a nontrivial Berry phase winding and therefore, also persists in systems where angular momentum is not conserved~\cite{Liu2020}. Together with the fact that our numerical studies were carried out in a photonic analog of an atomic borophene allotrope, this paves the way of harnessing the unique properties of pseudospin-2 conical intersections in photonic applications such as pseudospin coupling and the generation of nano-scale higher-charged optical vortices~\cite{ni2021multidimensional}, or scale-invariant lasing~\cite{contractor2022scalable}. Furthermore, the existence of two chiral variants of our borophene lattice combined with their pseudospin-2 conical intersections could provide additional interesting opportunities, e.g., in bilayer borophene stacking~\cite{CrastodeLima2019a,Albuhairan2022} or chiral topological photonics~\cite{mehrabad2020chiral}.
\appendix
\renewcommand\thesection{APPENDIX~\Alph{section}}
\renewcommand\thesubsection{\Alph{section}\arabic{subsection}}
\section{METHODS}\label{sec:methods}
\subsection{Numerically simulated beam propagation in the lattice}
The $z$-propagation of a slowly varying envelope light field $\psi(x,y,z)$ through a photonic lattice in the paraxial approximation is well described by the following continuous-model Schrödinger-type equation~\cite{Longhi2009}
\begin{multline}\label{eq:NLSE}
    \text{i}\frac{\partial }{\partial z} \psi(x,y,z) = \left [ -\frac{1}{2k_0n_0}\nabla_\bot^2 - k_0\Delta n(x,y) \right ] \psi(x,y,z) \\
    = \hat{H} \psi(x,y,z).
\end{multline}
Here $n_0$ is the background refractive index, $k_0$ is the wave number in vacuum, $\nabla_\bot=(\partial_x,\partial_y)$, and $\Delta n(x,y)$ represents the transverse refractive index change of the photonic lattice. For a sufficiently small transverse refractive index modulation, solutions to Eq.~\ref{eq:NLSE} can be obtained via the split-step Fourier transform method~\cite{Sharma2004a}. In our numerical simulations we adapt the parameters to match previously reported experiments in laser-written photonic lattices~\cite{hanafi2022localized,hanafi2022b}. We choose $n_0=1.4$ and $k_0 = 2\pi/\lambda_0$ with $\lambda_0=\SI{532}{\nano\metre}$. The photonic lattice potential consists of single-mode waveguides with a super-Gaussian refractive index potential and $\text{FWHM} = \SI{9}{\micro\metre}$, placed at a waveguide separation of $\Lambda = \SI{18}{\micro\metre}$ from each other. The selected potential strength is $\Delta n = 1.3 \times 10^{-3}$.

For single-mode evanescently coupled waveguides, the tight-binding approximation is valid, and the continuous Schrödinger Equation can be replaced by a discrete version. From the associated Hamiltonian for nearest neighbors only, which is given in k-space by Eq.~\ref{eq:Ham_snub_boro}, we can calculate the band structure. To confirm that the tight-binding approximation well-describes the dynamics in the chiral borophene lattice, we can solve the discrete version of Eq.~\ref{eq:NLSE} and compare the results with the continuous model. We do so by numerically solving the $N$ coupled differential equations resulting from the real space Hamiltonian of a chiral borophene lattice of $N$ lattice sites. The solutions obtained via the \textit{ode45} function of MATLAB perfectly match those of the continuous model (see Fig.~\ref{fig:comparison}).
\subsection{Pseudospin filtering}
To project the conically diffracted output light fields onto the different pseudospin eigenstates, we consider the tight-binding limit. As we need to assign one complex amplitude value to each waveguide of the continuous model we average over the area of the waveguide in the output light fields. We apply this averaging to each lattice site A to F and, for every unit cell at $\mathbf{R}=n\mathbf{a_1}+m\mathbf{a_2}$, we obtain a six-dimensional state vector $\ket{\psi_{\mathbf{R}}^{\text{out}}}$. We then calculate the projections of this vector onto the five pseudospin eigenstates as $\braket{\psi_{\mathbf{R}}^{\text{out}}|\psi_{-2,-1,0,+1,+2}}$, obtaining five complex amplitudes for each unit cell.

\section{Intuitive derivation of the pseudospin eigenstates -1, 0, +1, +2}\label{sec:deriv}
\begin{figure}
\centering\includegraphics[width=\linewidth]{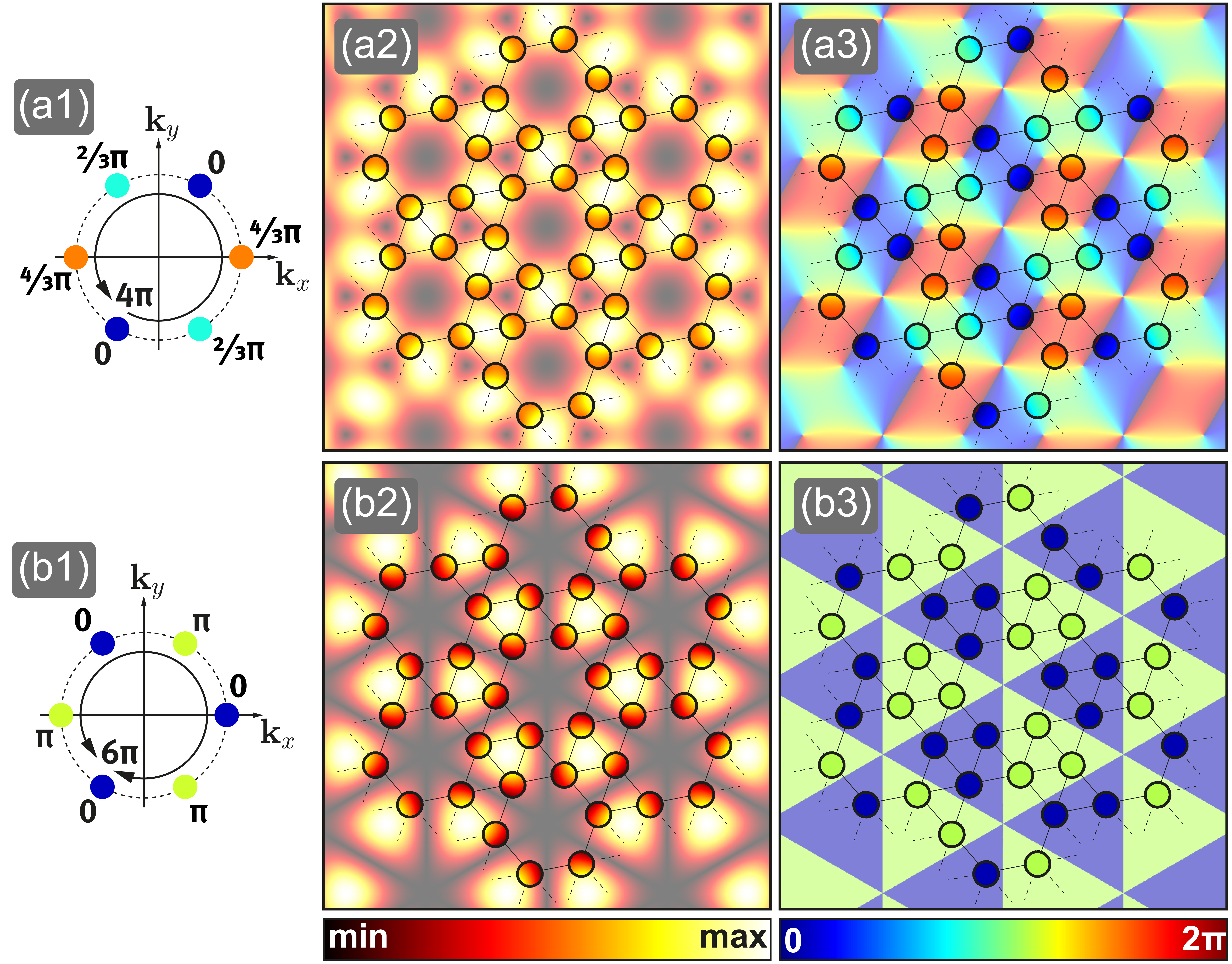} 
\caption{Derivation of the pseudospin eigenstates $\ket{\psi_{-1}}$ and $\ket{\psi_{0}}$. (a1) Six plane waves in k-space forming a hexagonal discrete vortex with $l=+2$. (a2) and (a3) Real space transverse amplitude and phase profile of the resulting discrete nondiffracting beam with overlaid chiral borophene lattice for $\ket{\psi_{-1}}$. (b1) Same as in (a1) but with with $l=\pm3$. (b2) and (b3) Same as in (a2) and (a3) but for $\ket{\psi_{0}}$.}
\label{fig:ps_derivation}
\end{figure}

\begin{figure}
\centering\includegraphics[width=\linewidth]{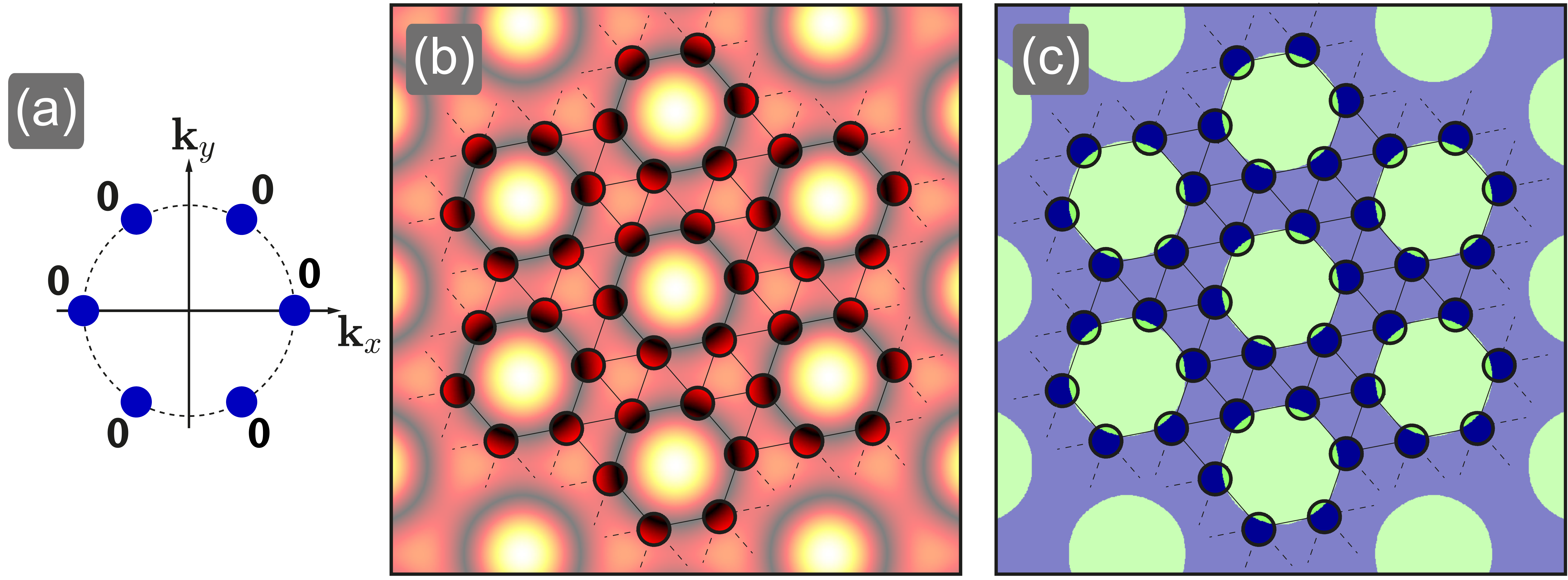} 
\caption{Low index mode. (a) Six plane waves in k-space without discrete vortex. (b) and (c) Real space transverse amplitude and phase profile of the resulting discrete nondiffracting beam with overlaid chiral borophene lattice.}
\label{fig:low_index}
\end{figure}

As stated in the main article, we present an intuitive approach to derive the pseudospin eigenstates by looking at the family of nondiffracting beams resulting  from the interference of six plane waves. The derivation of two of the four remaining eigenstates obtained by different phase relations of the interfering plane waves is summarized in Fig.~\ref{fig:ps_derivation}.
For the remaining eigenstates $\ket{\psi_{+1}}$ and $\ket{\psi_{+2}}$, the results are analogous to $\ket{\psi_{-1}}$ and $\ket{\psi_{-2}}$ albeit with opposing signs of vorticity in k- and in real-space. There is also a sixth nondiffracting beam which can be obtained by the interference of six in-phase plane waves. However, as shown in Fig.~\ref{fig:low_index}, this configuration leads to a low index mode in the photonic lattice where the light is concentrated in the unmodulated regions of the borophene lattice. This is not a tight-binding mode and, since it possesses a different propagation constant, it does not belong to the conical intersection.

\section{Effective Hamiltonian of the pseudospin-2 conical intersection}\label{sec:anal_deriv}
The effective Hamiltonian for pseudospin $s$ conical intersections can be expressed in terms of spin matrices of dimension $2s+1$ satisfying the angular momentum algebra $\left [ \hat{S}_i , \hat{S}_j  \right ] = \text{i}\varepsilon_{i,j,k}\hat{S}_k$. To show this is also the case for our chiral borophene lattice we Taylor expand the Hamiltonian of Eq.~\ref{eq:Ham_snub_boro} around the $\Gamma$-point to the first order. We obtain
\begin{equation}
    \hat{H}_\mathbf{k}{=}t \begingroup \setlength\arraycolsep{3pt} \begin{pmatrix}
0 & 1 & 1{-}\text{i}\mathbf{a}_1\mathbf{k} & 1{-}\text{i}\mathbf{a}_2\mathbf{k} & 1{-}\text{i}\mathbf{a}_2\mathbf{k} & 1 \\
1 & 0 & 1 & 1{-}\text{i}\mathbf{a}_2\mathbf{k} & 1{+}\text{i}\mathbf{a}_3\mathbf{k} & 1{+}\text{i}\mathbf{a}_3\mathbf{k} \\
1{+}\text{i}\mathbf{a}_1\mathbf{k} & 1 & 0 & 1 & 1{+}\text{i}\mathbf{a}_3\mathbf{k} & 1{+}\text{i}\mathbf{a}_1\mathbf{k}\\
1{+}\text{i}\mathbf{a}_2\mathbf{k} & 1{+}\text{i}\mathbf{a}_2\mathbf{k} & 1 & 0 & 1 & 1{+}\text{i}\mathbf{a}_1\mathbf{k}\\
1{+}\text{i}\mathbf{a}_2\mathbf{k} & 1{-}\text{i}\mathbf{a}_3\mathbf{k} & 1{-}\text{i}\mathbf{a}_3\mathbf{k} & 1 & 0 & 1\\
1 & 1{-}\text{i}\mathbf{a}_3\mathbf{k} & 1{-}\text{i}\mathbf{a}_1\mathbf{k} & 1{-}\text{i}\mathbf{a}_1\mathbf{k} & 1 & 0
\end{pmatrix}\endgroup,
\end{equation}
with $\mathbf{k}=(k_x,k_y)$, $\mathbf{a_1} = d/2(\sqrt{3},1)$, $\mathbf{a_2} = d(0,1)$, and $\mathbf{a_3} = \mathbf{a_1} - \mathbf{a_2}$. We can change the basis from the sublattice one to the orbital angular momentum basis by calculating $\hat{H}'=U^{\dagger}\hat{H}U$, with the unitary matrix $U$ composed of the normalized pseudospin eigenstates that have been appropriately phase-shifted
\begin{equation}\label{eq:unit_matrix}
    U = \frac{1}{\sqrt{6}} \begin{pmatrix}
e^{\text{i}\frac{5}{3}\pi} & e^{\text{i}\frac{5}{3}\pi} & 1 & e^{\text{i}\frac{4}{3}\pi} & e^{\text{i}\frac{1}{3}\pi} & 1\\
e^{\text{i}\frac{4}{3}\pi} & -1 & -1 & 1 & e^{\text{i}\frac{2}{3}\pi} & 1\\
-1 & e^{\text{i}\frac{1}{3}\pi} & 1 & e^{\text{i}\frac{2}{3}\pi} & -1 & 1\\
e^{\text{i}\frac{2}{3}\pi} & e^{\text{i}\frac{5}{3}\pi} & -1 & e^{\text{i}\frac{4}{3}\pi} & e^{\text{i}\frac{4}{3}\pi} & 1\\
e^{\text{i}\frac{\pi}{3}} & -1 & 1 & 1 & e^{\text{i}\frac{5}{3}\pi} & 1\\
1 & e^{\text{i}\frac{\pi}{3}} & -1 & e^{\text{i}\frac{2}{3}\pi} & 1 & 1
\end{pmatrix}.
\end{equation}
Discarding the state $\ket{\psi_{\beta=5}}$ of the sixth band by eliminating the last row and column, we obtain the effective 5x5 Hamiltonian
\begin{equation}\label{eq:Ham_eff}
    \hat{H}_{\text{eff}}{=}t \begingroup \setlength\arraycolsep{0pt} \begin{pmatrix}
-1 & \frac{1}{2}(k_x{-}\text{i}k_y) & 0 & 0 & 0 \\
\frac{1}{2}(k_x{+}\text{i}k_y) & -1 & \frac{1}{2}(k_x{-}\text{i}k_y) & 0 & 0 \\
0 & \frac{1}{2}(k_x{+}\text{i}k_y) & -1 & \frac{1}{2}(k_x{-}\text{i}k_y) & 0 \\
0 & 0 & \frac{1}{2}(k_x{+}\text{i}k_y) & -1 & \frac{1}{2}(k_x{-}\text{i}k_y) \\
0 & 0 & 0 & \frac{1}{2}(k_x{+}\text{i}k_y) & -1
\end{pmatrix}\endgroup,
\end{equation}
or in polar coordinates $\mathbf{k}=(k\cos{\theta},k\sin{\theta})$
\begin{equation}\label{eq:Ham_eff_polar}
    \hat{H}_{\text{eff}} = t \begin{pmatrix}
-1 & \frac{1}{2}ke^{-\text{i}\theta} & 0 & 0 \\
\frac{1}{2}ke^{\text{i}\theta} & -1 & \frac{1}{2}ke^{-\text{i}\theta} & 0 & 0 \\
0 & \frac{1}{2}ke^{\text{i}\theta} & -1 & \frac{1}{2}ke^{-\text{i}\theta} & 0 \\
0 & 0 & \frac{1}{2}ke^{\text{i}\theta} & -1 & \frac{1}{2}ke^{-\text{i}\theta} \\
0 & 0 & 0 & \frac{1}{2}ke^{\text{i}\theta} & -1
\end{pmatrix},
\end{equation}
whose eigenvalues are independent of the polar angle and have values of
\begin{equation}
    \beta_{1,2,3,4,5}{=}t({-}1{-}\sqrt{3}k),t({-}1{-}k),{-}t,t({-}1{+}k),t({-}1{+}\sqrt{3}k).
\end{equation}
The spectrum of the effective Hamiltonian is rotationally symmetric as expected from a conical intersection. For such a rotationally symmetric spectrum there is an associated conserved quantity. In this case, it is the $z$-component of the total angular momentum $J_z = S_z + L_z$. We can see this resulting from $[J_z,\hat{H}_{\text{eff}}]=[S_z,\hat{H}_{\text{eff}}]+[L_z,\hat{H}_{\text{eff}}]=0$, with $L_z = -\text{i}\partial/\partial\theta$ and the spin matrix
\begin{equation}
    S_z = \begin{pmatrix}
    2 & 0 & 0 & 0 & 0 \\
    0 & 1 & 0 & 0 & 0 \\
    0 & 0 & 0 & 0 & 0 \\
    0 & 0 & 0 & -1 & 0 \\
    0 & 0 & 0 & 0 & -2
    \end{pmatrix},
\end{equation}
The eigenvalues of $S_z$ correspond to the pseudospin values $m_s=+2,+1,0,-1,-2$ and, as can be seen by transforming the eigenstates back into the sublattice basis, its eigenstates are the pseudospin ones $\ket{\psi_{+2}},\ket{\psi_{+1}},\ket{\psi_{0}},\ket{\psi_{-1}},\ket{\psi_{-2}}$. As mentioned, the effective Hamiltonian can be expressed in terms of the spin matrices $\mathbf{S} = \left ( S_x,S_y \right )$ and $S_z$. With the standard spin-2 matrices
\begin{equation}
    S_x = \frac{1}{2}\begin{pmatrix}
    0 & 2 & 0 & 0 & 0 \\
    2 & 0 & \sqrt{6} & 0 & 0 \\
    0 & \sqrt{6} & 0 & \sqrt{6} & 0 \\
    0 & 0 & \sqrt{6} & 0 & 2 \\
    0 & 0 & 0 & 2 & 0
    \end{pmatrix},
\end{equation}
\begin{equation}
    S_y = \frac{1}{2\text{i}}\begin{pmatrix}
    0 & 2 & 0 & 0 & 0 \\
    -2 & 0 & \sqrt{6} & 0 & 0 \\
    0 & -\sqrt{6} & 0 & \sqrt{6} & 0 \\
    0 & 0 & -\sqrt{6} & 0 & 2 \\
    0 & 0 & 0 & -2 & 0
    \end{pmatrix},
\end{equation}
we get the following expression
\begin{equation}\label{eq:Heff_S}
    \hat{H}_{\text{eff}}(\mathbf{k}) = c_0\mathbf{k} \cdot \mathbf{S} + c_1\mathbf{k} \cdot \left \{ \mathbf{S},S_z^2 \right \} - t I_5,
\end{equation}
with the anticommutator $\left \{ \mathbf{S},S_z^2 \right \} = \mathbf{S} S_z^2 + S_z^2 \mathbf{S}$, $c_0 = \frac{1}{24}t(5\sqrt{6}-3)$, $c_1 = \frac{1}{24}t(3-\sqrt{6})$, and the identity matrix $I_5$ representing a shift of the conical intersection towards $-t$. The $\left \{ \mathbf{S},S_z^2 \right \}$ term in Eq.~\ref{eq:Heff_S} allows the control of the relative opening angle of the two pairs of cones. While in a standard s-2 conical intersection the slope of the inner cone is double that of the outer cone, in our spectrum there is a factor of $\sqrt{3}$.

By introducing $S_z$ raising and lowering operators $S_{\pm}=S_x \pm S_y$, we can recast Eq.~\ref{eq:Heff_S} as
\begin{multline}\label{eq:Heff_ladder}
    \hat{H}_{\text{eff}}(k,\theta) = \frac{c_0k}{2} \left [ e^{-\text{i}\theta}S_+ + e^{\text{i}\theta}S_-  \right ] \\
    + \frac{c_1k}{2} \cdot \left [ e^{-\text{i}\theta}\left \{ S_+,S_z^2 \right \} + e^{\text{i}\theta}\left \{ S_-,S_z^2 \right \} \right ] - t I_5.
\end{multline}
From Eq.~\ref{eq:Heff_ladder} we can read that by applying the Hamiltonian, which equates to propagation in our photonic lattice, $S_z$ is raised (lowered) while a phase factor of $e^{-\text{i}\theta}$ ($e^{\text{i}\theta}$) is introduced. This phase factor accounts for the negative (positive) optical phase vortices that are created during the propagation of a conically diffracting pseudospin ($S_z$) eigenstate which couples to the other eigenstates.

\section{Comparison between simulations in the continuous model and in the tight-binding model}\label{sec:comparison}
\begin{figure}
\centering\includegraphics[width=\linewidth]{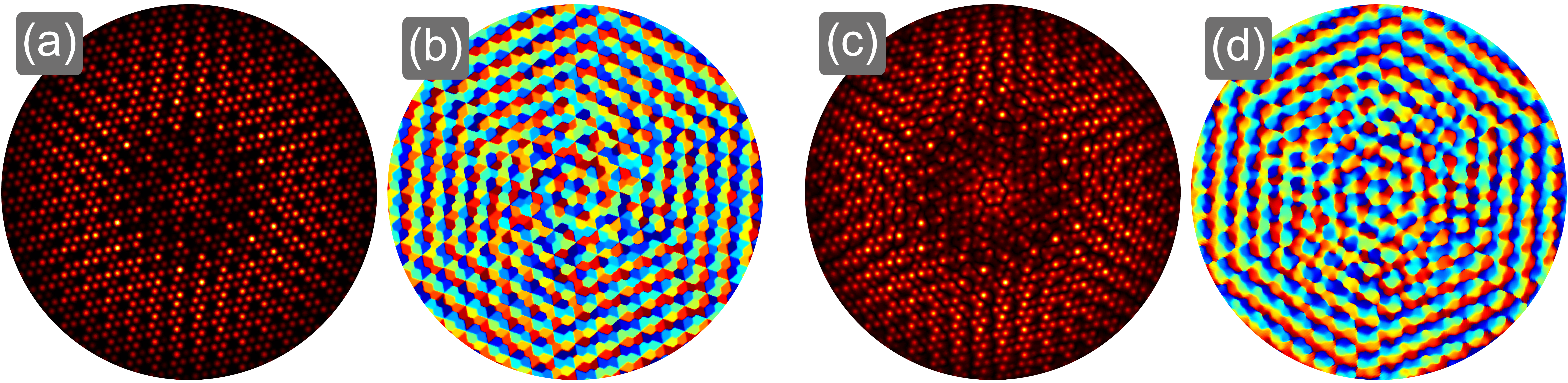} 
\caption{Comparison of numerical simulations in tight-binding and continuous model. (a)-(b) Amplitude and phase from solving the tight-binding coupled differential equations. (c)-(d) Amplitude and phase from solving the continuous model via the split-step beam propagation method.}
\label{fig:comparison}
\end{figure}

We want to show that the tight-binding model with only nearest neighbor coupling describes our photonic lattice well. To this aim, we compare simulations obtained by solving the continuous Schrödinger equation via the split-step method with those obtained by solving the $N$ coupled differential equations for a lattice of $N$ waveguides. We excite the lattice with the eigenstate $\ket{\psi_{-2}}$ with a Gaussian envelope of $\text{FWHM}=\SI{120}{\micro\metre}$. After propagation in the chiral borophene lattice for a propagation distance of four coupling lengths $L_c$, the output profiles obtained by the two numerical methods, as shown in Fig.~\ref{fig:comparison} match extremely well.

\section{Conical diffraction and topological charge conversion of the pseudospin eigenstates -1, 0, +1, +2}\label{sec:remaining}
\begin{figure}
\centering\includegraphics[width=\linewidth]{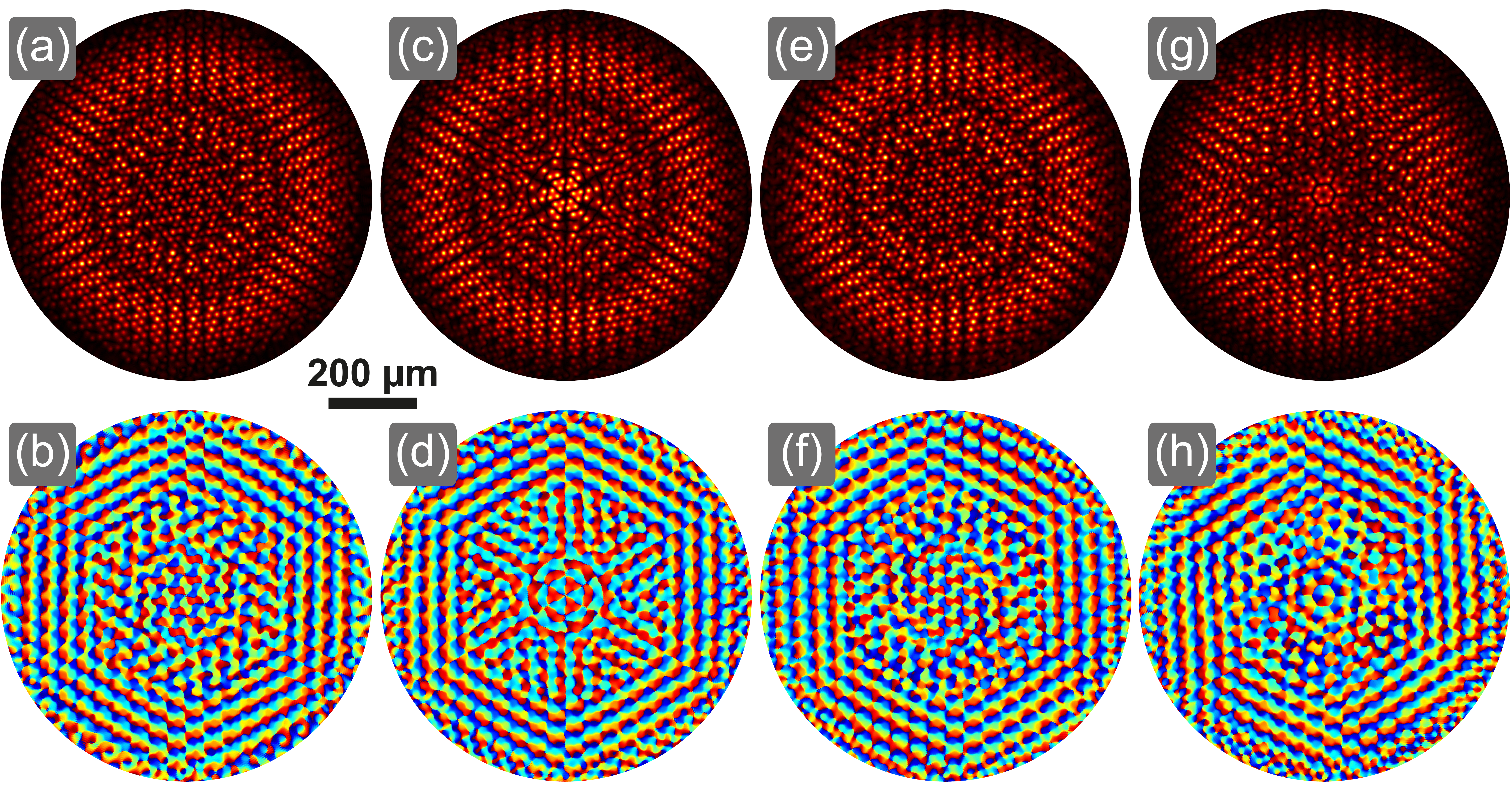} 
\caption{Numerically simulated conical diffraction outputs of the remaining pseudospin eigenstates. (a)-(b) Amplitude and phase after propagation of $\ket{\psi_{-1}}$ in the chiral borophene lattice. (c)-(d) Same as (a)-(b), but for $\ket{\psi_{0}}$ as input. (e)-(f) Same as (a)-(b), but for $\ket{\psi_{+1}}$ as input. (g)-(h) Same as (a)-(b), but for $\ket{\psi_{+2}}$ as input.}
\label{fig:rest_simul}
\end{figure}

The observation of conical diffraction and generation of optical phase vortices via topological charge conversion between different pseudospin eigenstates provided in the main part is already a complete demonstration of our pseudospin-2 conical intersection. However, to complete the picture, we repeat the procedure for all pseudospin eigenstates. The results obtained after beam propagation simulations of the other four pseudospin eigenstates are shown in Fig.~\ref{fig:rest_simul}. All parameters for the simulations were kept the same as in previous part. We can see that all states diffract conically, although not with the same expansion rate. This is to be expected since wave packets with larger pseudospin values expand more slowly~\cite{Diebel2016b}. The output light fields obtained using respectively $\ket{\psi_{-1}}$ and $\ket{\psi_{+1}}$ as the inputs are essentially equivalent except for the inverted phase vorticity. The same goes for those obtained respectively from $\ket{\psi_{-2}}$ and $\ket{\psi_{+2}}$.

\begin{figure}
\centering\includegraphics[width=\linewidth]{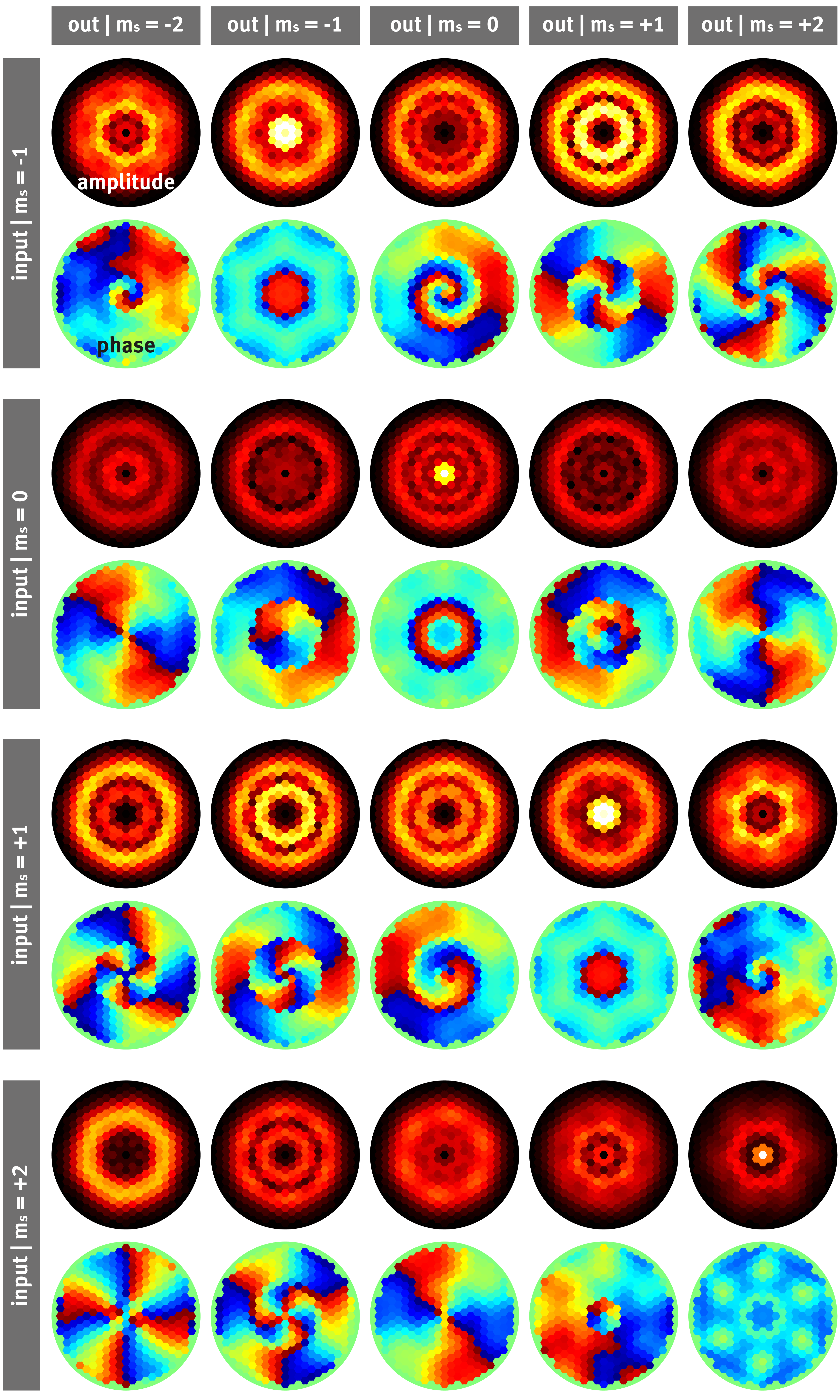} 
\caption{Decompositions of the conical diffraction outputs of the remaining pseudospin eigenstates. The amplitudes for different input states are scaled to the maximum of the corresponding row. The optical phase vortices have topological charge obeying the relation $l=m_s^{in}-m_s^{out}$.}
\label{fig:rest_decompo}
\end{figure}

We then proceed and decompose the output fields of Fig.~\ref{fig:rest_simul} by calculating the projection of them onto the different pseudospin eigenstates unit cell by unit cell. We obtain five projections for each of the four light fields. All 20 projections are shown in Fig.~\ref{fig:rest_decompo}. We can see that the total angular momentum $J_z$ is conserved in all cases via the generation of optical phase vortices having topological charge obeying the relation $l=m_s^{in}-m_s^{out}$.

\begin{backmatter}

% \bmsection{Acknowledgments} P.M., H.H., J.I., and C.D. gratefully acknowledge support from the Open Access Publication Fund of the University of Münster.

% \bmsection{Disclosures} The authors declare no conflicts of interest.

% \bmsection{Data availability} Data underlying the results presented in this paper are not publicly available at this time but may be obtained from the authors upon reasonable request.

\end{backmatter}

% Bibliography
\bibliography{sample}

% Full bibliography added automatically for Optics Letters submissions; the following line will simply be ignored if submitting to other journals.
% Note that this extra page will not count against page length
\bibliographyfullrefs{sample}

%Manual citation list
%\begin{thebibliography}{1}
%\bibitem{Zhang:14}
%Y.~Zhang, S.~Qiao, L.~Sun, Q.~W. Shi, W.~Huang, %L.~Li, and Z.~Yang,
 % \enquote{Photoinduced active terahertz metamaterials with nanostructured
  %vanadium dioxide film deposited by sol-gel method,} Opt. Express \textbf{22},
  %11070--11078 (2014).
%\end{thebibliography}

\end{document}